\documentclass[11pt,a4paper,english,nofootinbib,superscriptaddress]{revtex4}
\usepackage{lmodern}

\usepackage[T1]{fontenc}
\usepackage{babel}
\usepackage{color}
\usepackage{cancel}
\usepackage{amsmath}
\usepackage{graphicx}
\usepackage{amssymb}
\usepackage{esint}
\usepackage[unicode=true, pdfusetitle,
bookmarks=true,bookmarksnumbered=false,bookmarksopen=false,
breaklinks=false,pdfborder={0 0 1},backref=false,colorlinks=false]
{hyperref}
\usepackage{amsmath}
\usepackage{amssymb}
\usepackage[compat=1.1.0]{tikz-feynman}
\usepackage{float}

\def\barra#1{\not \!#1}
\def\b{\begin{equation}} \def\e{\end{equation}}
\def\bd{\begin{displaystyle}} \def\ed{\end{displaystyle}}
\def\ba{\begin{array}} \def\ea{\end{array}}

\def\bee{\begin{enumerate}}
	\def\eee{\end{enumerate}}

\def\le{\langle}
\def\re{\rangle}

\def\ud{\mathrm{d}}
\def\le{\langle}
\def\re{\rangle}
\def\dg{^{\dag}}

\def\1{\mbox{I\hspace{-.15em}1}}

\def\R{{\rm I\hspace{-.15em}R}}

\def\b{\begin{equation}}
\def\e{\end{equation}}
\def\bee{\begin{enumerate}}
	\def\eee{\end{enumerate}}

\makeatletter
\usepackage{latexsym}\usepackage{bm}

\makeatother
\begin{document}
	
\title{Photon decaying in de Sitter universe}
	
\author{Y. Ahmadi}
\email{ahmadi.pave@gmail.com} \affiliation{Department of
Physics, Razi University, Kermanshah, Iran}
	
\author{M.V. Takook \footnote{{corresponding author: M. V. Takook}}}
\email{ takook@razi.ac.ir} \affiliation{Department of Physics, Kharazmi University, Tehran, Iran}

\begin{abstract}

The interaction between three photons is studied in de Sitter ambient space formalism. As a special case the half harmonic generator is considered, {\it i.e.} one photon decays to two same-energy photons. The scattering matrix elements are presented which define the indirect gravitational effect on quantum field theory. The null curvature limit of scattering matrix is obtained for comparing it with its Minkowskian counterpart. The Hamiltonian of this interaction, in Minkowski space-time, was presented by using the quantum vacuum fluctuation in the one-loop approximation.

\end{abstract}
\maketitle

\section{introduction}

Historically, the interaction of light with matter is one of the most important interactions in physics. This interaction plays an important role in technology, which can be explained completely in quantum field theory (QFT) model.  In quantum level the interaction between photons, {\it i.e.} wave mixing, has a special role in quantum optics such as fiber Laser and squeezed light. The process of decaying one photon to two photons named second  harmonic generator (SHG). The reverse of this process {\it i.e.} merging tow photons and exiting one photon is also possible. These interactions are performed by using some crystals. In flat space, SHG is used for generating the coherent state \cite{karsten,garrison,fox,walls}. In quantum optics, the Hamiltonian of SHG process can be written as ${\cal H}= {\hslash}{\chi}(d^{\dagger}\;d'\;d''+d\;d'^{\dagger}\;d''^{\dagger})$ \cite{walls,shg}. The $d$ and $d^\dagger$ are annihilation and creation operators respectively, and $\chi$ is interaction constant which fix phenomenologically \cite{grynberg,karsten,bachor,fox,walls}. $\chi$ can be obtained from QFT point of view, that it done in this paper in \ref{flat int}.

In curved space-time, the interaction of fields is affected of background gravitational field. By computing the scattering matrix this indirect gravitational effects can be seen. In previous papers the interaction of vector and spinor fields and also the interaction of scalar and spinor fields are considered and the scattering matrix is calculated \cite{jaahta2017,ahjata2019}. By attention to that the experimental data clear that the our universe and early universe can be described by de Sitter (dS) space-time \cite{Riess, Perl,Nature,Henry 1,Henry 2,BICEP2} the construction of quantum field theory (QFT) in dS space-time is important for quantum gravity and unified theory. Some efforts have been made for constructing QFT in dS ambient space formalism in the last few years \cite{ta97,77,ta96,taazba,taro12,berotata,derotata,tami2012,meja2015,de2012,fa2015}. One of the advantages of ambient space formalism is the simplicity in calculations of probability amplitude because the action of dS group on ambient space coordinate is linear \cite{77}. In this paper, the decaying of one photon to two same-energy photons in dS ambient space formalism is investigated and the scattering matrix of this interaction is obtained in the first approximation. Finally its null curvature limit is discussed.

The organization of this article is as follow. The notation and terminology that have been used in this article is presented in section \ref{notation}. The $\cal S$-matrix elements in the small curvature approximation is investigated in dS ambient space formalism in section \ref{ds int}. In section \ref{flat int}, the decaying of one photon to two same energy photons has been recalled in Minkowski space-time. In section \ref{flat limit}, we calculate the null curvature limit of the $\cal S$-matrix elements and compare it with Minkowsian counterpart. Finally, the conclusion has been presented in section \ref{conclusion}.

\setcounter{equation}{0}
\section{notation}\label{notation}

The dS space-time can be considered as a four-dimensional hyperboloid embedded in five-dimensional Minkowski space-time:
$$
M_H=\left\{ x \in \R^5| \; \; x \cdot x=\eta_{\alpha\beta} x^\alpha
x^\beta =-H^{-2}\right\},\;\; \alpha,\beta=0,1,2,3,4\,.$$
The metric is:
\b \label{ds difiniion}
ds^2=\eta_{\alpha\beta}dx^{\alpha}dx^{\beta}|_{x^2=-H^{-2}}=g_{\mu\nu}^{dS}dX^{\mu}dX^{\nu},\;\; \mu,\nu=0,1,2,3,\e
where $\eta_{\alpha\beta}=$diag$(1,-1,-1,-1,-1)$, $H$ is Hubble constant parameter, $X^\mu$ is dS intrinsic coordinate and $x^\alpha$ is the five-dimensional dS ambient space formalism.

The action of dS group, $SO(1,4)$, on the intrinsic coordinate $X^\mu$ is complicated but the action of this group on the ambient space coordinate $x^\alpha$ is linear. So the calculation in this formalism is simple and similar to Minkowski space-time \cite{77}.
In this formalism the five matrices $\gamma^{\alpha}$ are needed which satisfy the following relations \cite{ta97,ta96,bagamota,77}:
\b \label{dS gamma relation}
\gamma^{\alpha}\gamma^{\beta}+\gamma^{\beta}\gamma^{\alpha}
=2\eta^{\alpha\beta}\;,\;
\gamma^{\alpha\dagger}=\gamma^{0}\gamma^{\alpha}\gamma^{0}\, ,\e
and they can be chosen as:
\b \label{gamma relation} 
\gamma^0=\left( \begin{array}{clcr} I & \;\;0 \\ 0 &-I \\ \end{array} \right)
,\;\;\;\gamma^4=\left( \begin{array}{clcr} 0 & I \\ -I &0 \\ \end{array} \right),\;
\gamma^1=\left( \begin{array}{clcr} 0 & i\sigma^1 \\ i\sigma^1 &0 \\
\end{array} \right)
,\;\;\gamma^2=\left( \begin{array}{clcr} 0 & -i\sigma^2 \\ -i\sigma^2 &0 \\
\end{array} \right)
, \;\;\gamma^3=\left( \begin{array}{clcr} 0 & i\sigma^3 \\ i\sigma^3 &0 \\
\end{array} \right)\, ,\e
where $\sigma^i$ $(i=1,2,3)$ are the Pauli matrices. The $\gamma^\alpha$ matrices in dS ambient space formalism are different from Minkowski $\gamma^{'\mu}$ matrices. The relation between them is \cite{bagamota}:
\b \label{minkofski gamma} \gamma^{'\mu}=\gamma^{\mu}\gamma^4.\e

The dS Dirac first-order field equation is \cite{77,ta97}:
\b \label{ds sp fi eq}
\left(\barra{x} \;\barra{\partial}^\top-2 \pm i\nu\right)\Psi(x)=0,\;\;\; \barra{x}=\gamma_\alpha x^\alpha=\gamma\cdot x\,,\e
where $\nu>0$ is related with dS mass parameter as $m_{f,\nu}^2=H^2(\nu^2+2\pm i\nu)$, and $\partial_{\alpha}^\top=\partial_{\alpha}+Hx_{\alpha}x.\partial$ is the transverse derivative.
The charged spinor field operator, which satisfy the field equation \eqref{ds sp fi eq}, is \cite{bagamota,77}:
\b \label{psi expansion}
\Psi(x)={\cal N}_p\int d\mu(\xi)\sum_{\sigma=\pm\frac{1}{2}}\left[a( \tilde{\xi},\sigma)(Hx.\xi)^{-2-i\nu} {\cal U} (x,\xi,\sigma)+b^{\dag}(\xi,\sigma)(Hx.\xi)^{-1+i\nu} {\cal V} (x,\xi,\sigma)\right],\e
where ${\cal N}_p$ is normalization constant and $\xi^\alpha=(\xi^0, \vec \xi, \xi^4)\in C^+=\left\lbrace \xi \in \R^5|\;\; \xi\cdot \xi=0,\;\; \xi^{0}>0 \right\rbrace$  is the transformed variable of $x^{\alpha}$ in positive cone. The $\xi^\alpha$ becomes the energy-momentum four-vector in null curvature limit. Also the $\tilde\xi$ is $\tilde{\xi}^\alpha=(\xi^0, -\vec \xi, \xi^4)$ and the $\ud \mu(\xi)$ is the $SO(4)$-invariant normalized volume. The explicit form of $\cal{U}$ and $\cal{V}$ were presented in \cite{bagamota}. $a^\dag (\xi,\sigma)$ and $b^\dag (\xi,\sigma)$ are creation operators which act on dS invariant vacuum sate $\lvert \Omega\re$ as:
\cite{77}:
$$
a^\dag (\xi,\sigma) \left| \Omega \right> \equiv\left|1_{\xi, \sigma}^{a} \right\rangle,\;\; b^\dag (\xi,\sigma) \left| \Omega \right> \equiv\left|1_{\xi, \sigma}^{b} \right\rangle\,.
$$
The anti-commutation relations for creation and annihilation operators are:
\b \label{anti-commutation relation}
\left\{a(\tilde{\xi}',\sigma') , a^\dagger(\xi,\sigma)\right\}=\delta^3(\xi-\xi')\delta_{\sigma \sigma'}
\;,\;\;\;\;\;\;
\left\{b(\tilde{\xi}',\sigma') , b^\dagger(\xi,\sigma)\right\}=\delta^3(\xi-\xi')\delta_{\sigma \sigma'}\,.
\e
Analytic field operator is defined in complex dS space-time as \cite{77,brmo}:
$$\Psi(x)=\lim_{y\rightarrow 0} \Psi(z)=\lim_{y\rightarrow 0} \Psi(x+iy). $$
The adjoint spinor $\bar{\Psi}(x)$ in ambient space formalism which defined as $\bar{\Psi}(x)=\Psi^\dagger(x)\gamma^0\gamma^4$,  satisfies this field equation  \cite{ta96,bagamota,ta97,77,morrota}:
\b\label{ad ds sp fi eq}  \bar{\Psi}(x)\gamma^4\left(\overleftarrow{\barra\partial}^\top{\barra x}-2\mp i\nu\right)=0\;.
\e
The analytic two-point function of spinor field is \cite{bagamota,77}:
\b \label{ds spinor two-point function}	\left<\Omega\right|\Psi_{i}(z_1)\bar \Psi_{j}(z_2)\left|\Omega\right>=i{\bf S}_{ij}(z_1,z_2)=\dfrac{c_{\frac{1}{2},\nu}}{2}\int_{S^3} d\mu(\xi) (z_1.\xi)^{-2-i\nu}(z_2.\xi)^{-2+i\nu}\left(\cancel{\xi}\gamma^4\right)_{ij},\e
where, $c_{\frac{1}{2},\nu}$ is the normalization constant. This analytic two-point function can be calculated in terms of generalized Legendre function of first kind \cite{ta96,ta97,bagamota}.

As it is discussed in \ref{flat limit}, the null curvature limit of the quantum field operator of spinor field $\Psi(x)$ reduces to its Minkowski counterpart $\psi(X)$.
In Minkoski space time the Dirac equation is $\left(i\barra\partial -m\right)\psi=0$. In this equation $\cancel{\partial}=\gamma^{'\mu}\;\partial_\mu$ and $\gamma^{'\mu}$ matrices are $4\times4$ matrices which satisfy the following conditions \cite{mandl shaw,kaku,zuber,ryder}:
\b\label{Minkowski gamma relation}\gamma^{'\mu}\gamma^{'\nu}+\gamma^{'\nu}\gamma^{'\mu}=2\eta^{\mu\nu}\;,\;\;\;\;\gamma^{'\mu\dagger}=\gamma^{'0}\gamma^{'\mu}\gamma^{'0},\e
that the $\eta_{\mu\nu}$ is Minkowski metric.

In the ambient space formalism one can write the massless vector field, similar to other massless quantum field, in terms of the massless conformally coupled scalar field. The field operator is obtained in terms of annihilation $d$ and creation $d^\dagger$ operators as \cite{gagarota,ta97}:
\b \label{vector field expressed} {\cal A}_\alpha(x)={\cal N}_k \int d\mu({\xi})_{B}\sum_{n} \left[d({\tilde{\xi}},n)(Hx.\xi)^{-2}{\cal E}_{2\alpha}(x,\xi,n)+d^{\dag}(\xi,n)(Hx.\xi)^{-1}{\cal E}_{1\alpha}(x,\xi,n)\right]\, ,\e
where the ${\cal N}_k$ is the normalization constant, $n=0,1,2,3$ are the polarization states and the $d\mu_{B}({\xi})$ is $d\mu({\xi})_B=2\pi^2r^3drd\mu({\xi})$ \cite{tak,ta97}. Also ${\cal E}_{2\alpha}$ and ${\cal E}_{1\alpha}$ are polarization vectors which studied in \cite{gagarota,ta97}. The commutation relation between creation and annihilation operators and their action on vacuum state $\lvert\Omega\re$ is \cite{gata,gagarota}:
$$
\left[d(\tilde{\xi}',n') , d^\dagger(\xi,n)\right]=\delta_{s^3}(\xi-\xi')\delta_{nn'} \;,\;\;\;\;\;\;d(\tilde\xi ,n)\lvert\Omega\re=0\;,\;\;\;\;\;\;d\dg(\xi ,n) \lvert\Omega\re=\lvert 1^d_{\xi ,n}\re
.$$
The vector field operator can be written as two ''positive'' and ''negative'' parts as:
\b\label{+,- part ds ve-fi}
{\cal A}_\alpha={\cal A}^{(+)}_\alpha+{\cal A}^{(-)}_\alpha
\;\;\;;\;\;\;\left\{
\begin{array}{clcr}
{\cal A}^{(+)}_\alpha(x)={\cal N}_k\displaystyle \int d\mu({\xi})_{B}\sum_{n} d({\tilde{\xi}},n)(Hx.\xi)^{-2}{\cal E}_{2\alpha}(x,\xi,n)\,,
\\
\\
\displaystyle
{\cal A}^{(-)}_\alpha(x)={\cal N}_k \int d\mu({\xi})_{B}\sum_{n} d^{\dag}(\xi,n)(Hx.\xi)^{-1}{\cal E}_{1\alpha}(x,\xi,n)\,.
\end{array}
\right.
\e
The null curvature limit of dS quantum operator of vector field in ambient space formalism is matched on Minkowski quantum vector field operator $A_\mu(X)$. 

\setcounter{equation}{0}
\section{Scattering matrix in $dS$ space-time}\label{ds int}

In ambient space formalism of dS space-time, the dS-Dirac field equation is invariant under $U(1)$ global symmetry,
but it is not invariant under the local $U(1)$ symmetry.
By changing the gauge covariant derivative $ D_\alpha=\partial^\top_\alpha +iq{\cal A}_\alpha,$ with derivative $\partial^\top_\alpha,$ one can obtain dS-Dirac local gauge invariant equation \cite{agt2005}. By applying this change 
in the free field electromagnetic Lagrangian, the interaction Lagrangian can be obtained as \cite{agt2005}:
\b \label{ds sp-ve int lagrangian}
 {\cal L}_0 =H\bar{\Psi}(x)\gamma^4(-i\barra{x}\; \barra{\partial}^{\top}+2i+\nu)\Psi(x),\;\;\; {\cal L}_{int} =qH\bar{\Psi}(x)\gamma^4 \cancel{x}\;\cancel{{\cal A}}(x)\Psi (x).\e
The $q$ in the null curvature limit can be considered as the electric charge $e$. As it is seen, one can write:
$$ {\cal H}_{int}=-{\cal L}_{int}=-qH\bar{\Psi}(x)\gamma^4 \cancel{x}\;\cancel{{\cal A}}(x)\Psi (x)\, .$$

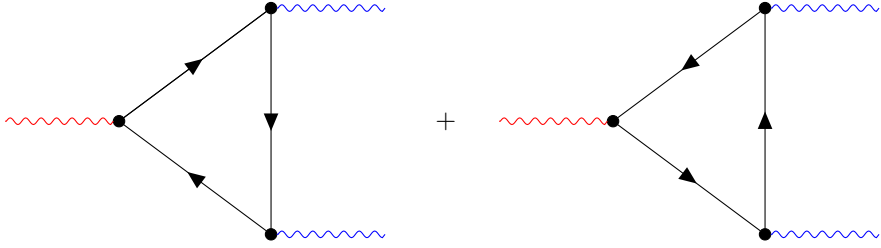
\begin{figure}[H]
	\centering
	\begin{tikzpicture}
	\begin{feynman}
	\vertex (a) at(0,0);
	\vertex [dot] (b) at(1.5,0) [] {};
	\vertex [dot] (h) at(3.5,1.5) [] {};
	\vertex [dot] (i) at(3.5,-1.5) [] {};
	\vertex (j) at(5,1.5);
	\vertex (k) at(5,-1.5);
	\vertex (e) at(6.5,0);
	\vertex [dot] (f) at(8,0)[] {}; 
	\vertex [dot] (l) at(10,1.5) [] {};
	\vertex (n) at(11.5,1.5) ;
	\vertex [dot] (m) at(10,-1.5)[] {};
	\vertex (o) at(11.5,-1.5);
	\vertex (+) at(5.8,0) {\(+\)};
	\diagram {
		(a) -- [photon,red] (b),
		(b) -- [fermion,dot] (h) --[fermion](i) --[fermion] (b),
		(h) -- [photon,blue] (j),
		(i) -- [photon,blue] (k),
	};	
	\diagram {
		(e) -- [photon,red] (f),
		(f)-- [fermion](m)-- [fermion](l)--[fermion](f),
		(l) -- [photon,blue] (n),
		(m) -- [photon,blue] (o),
	};	
	\end{feynman}
	\end{tikzpicture}
	\caption{interaction of three photons diagrams}
	\label{f}
\end{figure}

In dS space-time, the time evolution operator, $\vert \alpha , t\rangle=U(t,t_0) \vert \alpha,t_0 \rangle$ is only defined in the static coordinate system:
$$ x^\alpha\equiv \left(
\sqrt{H^{-2}-r^2}\sinh Ht_s\;,\;\sqrt{H^{-2}-r^2}\cosh Ht_s\;,\;r\cos\theta\;,\;r \sin \theta\cos\phi\;,\;r\sin\theta\sin \phi
\right), $$
where $-\infty<t_s<\infty\;,\;0\leq r<H^{-1}\;,\;0\leq \theta\leq\pi\;,\;0\leq \phi< 2\pi$. This coordinate system does not cover all dS hyperboloid. The time evolution operator for a quantum dS black-hole was considered in this coordinate system \cite{ta17}. But generally, in curved space-time, it is very complicated or impossible to calculate the $\cal S$-matrix elements in dS space due to the event horizon \cite{flilto,anilto,ta17}. In this work, because of occurring the interaction in atomic dimension, we can ignore the direct effect of curvature in this dimension but the indirect effect of curvature exists. In this approximation, the time evolution operator can be expanded in terms of Minkowskian counterpart:
$$U(t,t_0)={U_M}(t,t_0)+ Hf(t,t_0)+...\;.$$
The $ U_M$, in the null curvature limit, is exactly the Minkowski time evolution operator but the term $f(t,t_0)$ is due to the direct effect of curvature. Given that, the interaction is in atomic level, one can ignore the $f(t,t_0)$.
Therefore, although it is very difficult to calculate the scattering matrix in curved space, in atomic interaction approximation, the ${\cal S}$ matrix can be presented in the following approximate equation to calculate the indirect effect of curvature corrections from the Minkowski counterpart \cite{jaahta2017,ahjata2019}:
\b \label{ds s matrix expansion}
	\displaystyle{\cal S}\simeq \sum_{\lambda=0}^{\infty}{\cal S}^{(\lambda)},\;\;
	\displaystyle{\cal S}^{(\lambda)}=\dfrac{(-i)^\lambda}{\lambda!}\int\ud^4x_1\cdots \int\ud^4x_\lambda\;\;T\left[{\cal H}_{int}(x_1)\;\cdots\;{\cal H}_{int}(x_\lambda)\right].
\e
Therefore, by inserting \eqref{ds sp-ve int lagrangian} in \eqref{ds s matrix expansion} for $\lambda=3$ and by writing the matrices in terms of their components, the ${\cal S}^{(3)}$ is:
 \b\label{ds sp-ve int s3 element}
\begin{array}{clcr}
	{\cal S}^{(3)}&=\dfrac{i}{6\hslash^3}\displaystyle\int\int\int d\mu(x_1) d\mu(x_2)d\mu(x_3)T\left[{\cal H}(x_1){\cal H}(x_2){\cal H}(x_3)\right]
	\\
	\\
	&=-\dfrac{iq^3}{6\hslash^3}\displaystyle\int\int\int d\mu(x_1) d\mu(x_2)d\mu(x_3)\;\;T\left[
	\bar{\Psi}_i(x_1)\left(\gamma^4\;H\cancel{x}_1\right)_{il}\;\cancel{{\cal A}}_{lj}(x_1)\Psi_j(x_1)
	\right.
	\\
	\\
    &\times \left.
	\bar{\Psi}_{i'}(x_2)\left(\gamma^4\;H\cancel{x}_2\right)_{i'l'}\;\cancel{{\cal A}}_{l'j'}(x_2)\Psi_j'(x_2)
	\;\;
	\bar{\Psi}_{i''}(x_3)\left(\gamma^4\;H\cancel{x}_3\right)_{i''l''}\;\cancel{{\cal A}}_{l''j''}(x_3)\Psi_{j''}(x_3)
	\right].
\end{array}
\e
By using Wick theorem, one can write the time order product in terms of normal order product. The time order product becomes to 76 normal order products, but only two terms describe interactions $\gamma\rightarrow\gamma'+\gamma''$ and $\gamma'+\gamma''\rightarrow\gamma$. Therefore by notice to \eqref{ds spinor two-point function}, the 
${\cal S}^{(3)}_{(\gamma\rightarrow\gamma'+\gamma''\;,\;\gamma'+\gamma''\rightarrow\gamma)}$ is:
\b\label{ds sp-ve int s3 element 2}
\begin{array}{clr}
{\cal S}^{(3)}_{(\gamma\rightarrow\gamma'+\gamma''\;,\;\gamma'+\gamma''\rightarrow\gamma)}=-\dfrac{q^3}{6\hslash^3}\displaystyle\int\int\int d\mu(x_1) d\mu(x_2)d\mu(x_3)
N\left[
\cancel{{\cal A}}_{lj}(x_1)
\cancel{{\cal A}}_{l'j'}(x_2)
\cancel{{\cal A}}_{l''j''}(x_3)
\right]
\\
\\
\times\left\{
\left(\gamma^4H\cancel{x}_1\right)_{il}
{\bf S}_{ji'}(x_1,x_2)
\left(\gamma^4H\cancel{x}_2\right)_{i'l'}
{\bf S}_{j'i''}(x_2,x_3)
\left(\gamma^4H\cancel{x}_3\right)_{i''l''}
{\bf S}_{j''i}(x_3,x_1)
\right.
\\
\\
+
\left.
\left(\gamma^4H\cancel{x}_1\right)_{il}
{\bf S}_{ji''}(x_1,x_3)
\left(\gamma^4H\cancel{x}_3\right)_{i''l''}
{\bf S}_{j''i'}(x_3,x_2)
\left(\gamma^4H\cancel{x}_2\right)_{i'l'}
{\bf S}_{j'i}(x_2,x_1)
\right\},
\end{array}
\e
two terms of \eqref{ds sp-ve int s3 element 2} describe bellow diagrams with counterclockwise and anticlockwise spinor loops as shown in bellow figure \ref{f}.

The second term in \eqref{ds sp-ve int s3 element 2} by exchanging $x_1$ and $x_2$ becomes to first term, so one can consider one of them. Therefore:
$$
\begin{array}{cl}
{\cal S}^{(3)}_{(\gamma\rightarrow\gamma'+\gamma''\;,\;\gamma'+\gamma''\rightarrow\gamma)}=-\dfrac{q^3}{3\hslash^3}
\displaystyle\int\int\int d\mu(x_1) d\mu(x_2)d\mu(x_3)
N\left[
\cancel{{\cal A}}_{lj}(x_1)
\cancel{{\cal A}}_{l'j'}(x_2)
\cancel{{\cal A}}_{l''j''}(x_3)
\right]
\\
\\
\times\left\{
\left(\gamma^4H\cancel{x}_1\right)_{il}
{\bf S}_{ji'}(x_1,x_2)
\left(\gamma^4H\cancel{x}_2\right)_{i'l'}
{\bf S}_{j'i''}(x_2,x_3)
\left(\gamma^4H\cancel{x}_3\right)_{i''l''}
{\bf S}_{j''i}(x_3,x_1)
\right\}.
\end{array}
$$
By separating the $\cancel{\cal A}$ to positive and negative parts \eqref{+,- part ds ve-fi}, one can obtain the $N\left[
\cancel{{\cal A}}_{lj}(x_1)
\cancel{{\cal A}}_{l'j'}(x_2)
\cancel{{\cal A}}_{l''j''}(x_3)
\right]$ as:

\b \label{ds physical terms}
\begin{array}{clcr}
N\left[
\cancel{{\cal A}}_{lj}(x_1)\cancel{{\cal A}}_{l'j'}(x_2)\cancel{{\cal A}}_{l''j''}(x_3)
\right]
=\cancel{{\cal A}}^{(+)}_{lj}(x_1)\cancel{{\cal A}}^{(+)}_{l'j'}(x_2)\cancel{{\cal A}}^{(+)}_{l''j''}(x_3)
+\cancel{{\cal A}}^{(-)}_{lj}(x_1)\cancel{{\cal A}}^{(-)}_{l'j'}(x_2)\cancel{{\cal A}}^{(-)}_{l''j''}(x_3)
\\
\\
+\cancel{{\cal A}}^{(+)}_{lj}(x_1)\cancel{{\cal A}}^{(+)}_{l'j'}(x_2)\cancel{{\cal A}}^{(-)}_{l''j''}(x_3)+
\cancel{{\cal A}}^{(+)}_{lj}(x_1)\cancel{{\cal A}}^{(+)}_{l''j''}(x_3)\cancel{{\cal A}}^{(-)}_{l'j'}(x_2)+
\cancel{{\cal A}}^{(+)}_{l'j'}(x_2)\cancel{{\cal A}}^{(+)}_{l''j''}(x_3)\cancel{{\cal A}}^{(-)}_{lj}(x_1)
\\
\\
+\cancel{{\cal A}}^{(+)}_{lj}(x_1)\cancel{{\cal A}}^{(-)}_{l'j'}(x_2)\cancel{{\cal A}}^{(-)}_{l''j''}(x_3)+
\cancel{{\cal A}}^{(+)}_{l'j'}(x_2)\cancel{{\cal A}}^{(-)}_{lj}(x_1)\cancel{{\cal A}}^{(-)}_{l''j''}(x_3)+
\cancel{{\cal A}}^{(+)}_{l''j''}(x_3)\cancel{{\cal A}}^{(-)}_{lj}(x_1)\cancel{{\cal A}}^{(-)}_{l'j'}(x_2)
.
\end{array}
\e
Here, two first terms are unphysical and three terms in second line represent the interaction $\gamma'+\gamma''\longrightarrow\gamma$
and also three terms in third line represent  the interaction $\gamma\longrightarrow\gamma'+\gamma''$. Given that three terms in second line are equivalent, the contribution of them in calculations are the same. Thus by using \eqref{+,- part ds ve-fi}, the ${\cal S}^{(3)}_{\gamma\longrightarrow\gamma'+\gamma''}$ is obtained as: 
 $$
{\cal S}^{(3)}_{(\gamma\rightarrow\gamma'+\gamma'')}=i\int\int\int\ud\mu(\xi_k)\ud\mu(\xi_{k'})\ud\mu(\xi_{k''})\sum_{r}\sum_{r'}\sum_{r''}
\;\;{\chi}^{dS}\;\;a^r(k) a^{\dagger r'}(k') a^{\dagger r''}(k'')\,,
$$
where
 \b\label{ds xi}
\begin{array}{cl}
	{\chi}^{dS}  
	&=\dfrac{iq^3}{\hslash^3}
	{\cal N}_k{\cal N}_{k'}{\cal N}_{k''}
	\displaystyle\int\int\int d\mu(x_1) d\mu(x_2)d\mu(x_3)
	\left(Hx_1\cdot\xi_k\right)^{-2}
	\left(Hx_2\cdot\xi_{k'}\right)^{-1}
	\left(Hx_3\cdot\xi_{k''}\right)^{-1}
	\\
	\\
	&
	\times 
	Tr
	\left\{
	\gamma^4\;H\cancel{x}_1\;\cancel{\cal E}_2\;
	{\bf S}(x_1,x_2)\;
	\gamma^4\;H\cancel{x}_2\;\cancel{\cal E}_1\;
	{\bf S}(x_2,x_3)\;
	\gamma^4\;H\cancel{x}_3\;\cancel{\cal E}_1\;
	{\bf S}(x_3,x_1)
	\right\}.
\end{array}
\e
In \eqref{ds xi}, one can write the ${\bf S}(x,x')$ as \eqref{ds spinor two-point function} or in terms of generalized Legendre function \cite{77} and then calculate the \eqref{ds xi} by numerical methods. These calculations may open the door to measuring the gravitational effects on quantum fields which can be measured in the laboratory scale of energy. We postpone this direct calculation to next works; and in this work we calculate only the null curvature limit of \eqref{ds xi} for comparing with Minkowski results.
\setcounter{equation}{0}
\section{Scattering matrix in null curvature limit }\label{flat int}

In flat space-time, the interaction Hamiltonian between the electromagnetic field and spinor field is ${\cal H}_{int}=e\bar{\psi}\cancel{A}\psi$ \cite{kaku,zuber,mandl shaw,ryder,wei}. The $\cal S$-matrix for $\lambda=3$  in Minkowski space is:
\b\label{s3}
\begin{array}{clcr}
	{\cal S}^{(3)} &=\dfrac{i}{3!}\displaystyle\int\ud^4X_1\;\int\ud^4X_2\;\int\ud^4X_3\;\;T\left[{\cal H}_{int}(X_1)\;{\cal H}_{int}(X_2)\;{\cal H}_{int}(X_3)\right]
	\\
	\\
	&=\dfrac{ie^3}{6}\displaystyle\int\ud^4x_1\;\int\ud^4X_2\;\int\ud^4X_3\;\;T\left[(\bar{\psi}\cancel{A}\psi)_{X_1}(\bar{\psi}\cancel{A}\psi)_{X_2}(\bar{\psi}\cancel{A}\psi)_{X_3}\right].
\end{array}\e
By using Wick theorem and definition of spinor two-point function \cite{mandl shaw}, similar to that came in section \ref{ds int}, the ${\cal S}^{(3)}$ for interaction of three photons is:
\b\label{flat s3}
\begin{array}{clcr}
	{\cal S}^{(3)}_{(\gamma\rightarrow\gamma' +\gamma''\;,\;\gamma'+\gamma''\rightarrow\gamma)}
	&={\cal S}^{(3)}_a+{\cal S}^{(3)}_b= -\dfrac{e^3}{6\hslash^3}\displaystyle\int\ud^4X_1\int\ud^4x_2\int\ud^4X_3
	\\
	\\
	&\times\left\{N\left[\cancel{A}(X_1)\;S(X_1,X_2)\;\cancel{A}(X_2)\;S(X_2,X_3)\;\cancel{A}(X_3)\;S(X_3,X_1)
	\right]\right.
	\\
	\\
	&+\left.
	N\left[\cancel{A}(X_1)\;S(X_1,X_3)\;\cancel{A}(X_3)\;S(X_3,X_2)\;\cancel{A}(X_2)\;S(X_2,X_1)
	\right]\right\}.
\end{array}
\e
As it be seen, these two terms are the same and similar to previous section one can consider only one of them. So by using vector field operator, spinor two-point function \cite{mandl shaw,kaku,ryder,wei,zuber} and delta function definition in flat space-time, after some mathematical calculations, the ${\cal S}^{(3)}$ for interaction  $\left(\gamma\longrightarrow\gamma'+\gamma''\right)$ obtains as follow:
$${\cal S}^{(3)}_{\left(\gamma\longrightarrow\gamma'+\gamma''\right)}=i\int\ud^3k\int\ud^3k'\int\ud^3k''\sum_{r}\sum_{r'}\sum_{r''} \;\;\chi \;\; d^r(k) d^{\dagger r'}(k') d^{\dagger r''}(k''),$$
where
\b\label{xi}
\begin{array}{c}
	{\chi} =-ie^3{\cal N}_k{\cal N}_{k'}{\cal N}_{k''}\;\displaystyle\int\ud^4p\int\ud^4p'\int\ud^4p''\;\delta^4(p'-p+k) \delta^4(p''-k'-p') \delta^4(p-p''-k'')
	\\
	\\
	\times\left\{\dfrac{Tr\left[(\barra{p}+m)\barra{\epsilon}^{(r)}(k)(\barra{p}'+m)\barra{\epsilon}^{(r')}(k')(\barra{p}''+m)\barra{\epsilon}^{(r'')}(k'')\right]}{(p^2-m^2+i\tau)(p'^2-m^2+i\tau)(p''^2-m^2+i\tau)}\right\}.
\end{array}
\e
Given that the trace of odd number of gamma matrices is zero, with respect to the Delta function property, one can obtain this relation:

\b\label{ctr1}
\begin{array}{clcr}
	{\chi} &=-ie^3{\cal N}_k{\cal N}_{k'}{\cal N}_{k''}\;\;\delta^4(k-k'-k'')\;\displaystyle\int\ud^4p \;\;\dfrac{1}{[p^2-m^2+i\tau][(p-k)^2-m^2+i\tau][(p-k'')^2-m^2+i\tau]}
	\\
	\\
	&\times\left\{
	Tr[{\barra p}{\barra{\epsilon}}{\barra p}{\barra{\epsilon}}^{\;'}{\barra p}{\barra{\epsilon}}^{\;''}]
	-Tr[{\barra p}{\barra{\epsilon}}{\barra p}{\barra{\epsilon}}^{\;'}{\barra k}^{''}{\barra{\epsilon}}^{\;''}]
	-Tr[{\barra p}{\barra{\epsilon}}{\barra k}{\barra{\epsilon}}^{\;'} \;{\barra p}{\barra{\epsilon}}^{\;''}]
	+Tr[{\barra p}{\barra{\epsilon}} {\barra k}{\barra{\epsilon}}^{\;'} {\barra k}^{''}{\barra{\epsilon}}^{\;''}]
	+m^2 Tr[{\barra p}{\barra{\epsilon}} {\barra{\epsilon}}^{\;'} {\barra{\epsilon}}^{\;''}]
	\right.
	\\
	\\
	&\left.+m^2 Tr[{\barra{\epsilon}} {\barra p}{\barra{\epsilon}}^{\;'} {\barra{\epsilon}}^{\;''}]
	+m^2 Tr[{\barra{\epsilon}} {\barra{\epsilon}}^{\;'} {\barra p}{\barra{\epsilon}}^{\;''}]
	-m^2 Tr[{\barra{\epsilon}} {\barra k} {\barra{\epsilon}}^{\;'}{\barra{\epsilon}}^{\;''}]
	-m^2 Tr[{\barra{\epsilon}} {\barra{\epsilon}}^{\;'} {\barra k}^{''}{\barra{\epsilon}}^{\;''}]
	\right\}\, ,
\end{array}
\e
where $\epsilon\equiv\epsilon^{(r)}\;,\;\epsilon'\equiv\epsilon^{(r')}\;,\;\epsilon''\equiv\epsilon^{(r'')}.\;$
The energy of outgoing photons are the same; and the energy of each one of them is equal to half energy of incoming photon's energy, $ k'_0=k''_0  =\frac{1}{2} k_0 $. By attention to $k^2=0$ and by supposing that the direction of outgoing photons are the same for simplicity, one can obtain:
\b\label{k'=k''}
\overrightarrow{k}'=\overrightarrow{k}''=\frac{1}{2} \overrightarrow{k},
\;\;\;\;\longrightarrow\;\;\;\;
k^{'\mu}=k^{''\mu}=\dfrac{1}{2}k^\mu\,.
\e
After some mathematical calculations, the constant $\chi$ is obtained as \cite{3photon}:

\b\label{chi constant2}
{\chi}=\dfrac{2}{3}\pi^2e^3\;(\gamma-1)\; \delta^4(k-k'-k'') \;{\cal N}_{k} {\cal N}_{k'} {\cal N}_{k''}
\left[
\left(k\cdot\epsilon'\right)\left(\epsilon\cdot\epsilon''\right)-\left(k\cdot\epsilon''\right)\left(\epsilon\cdot\epsilon'\right)
\right],
\e
where the $\gamma=0.5772157$ is Euler's constant and one can write $\gamma=\displaystyle-\Gamma(0)-\lim_{n \rightarrow 0}(n)$ \cite{refaei2013}. It is clear that the coefficient $ \chi$  is also depend on the energy-momentum and the polarization states of the photons.

\setcounter{equation}{0}
\section{null curvature limit}\label{flat limit}

 The dS manifold is a hyperboloid with radios $H^{-1}$. For large radius, or equivalently ${H\rightarrow0}$, the curvature of space-time vanishes so the ${H\rightarrow0}$ limit called null curvature limit. In this limit, dS space-time matches with Minkowski space-time. Following coordinate system is useful for calculating the null curvature limit:
  $$x^\alpha\equiv \left(H^{-1}\sinh(HX^0)\; , \dfrac{\overrightarrow{X}}{H\lVert\overrightarrow{X}\lVert}\cosh(HX^0)\sinh(H\lVert\overrightarrow{X}\lVert)\; , H^{-1}\cosh(HX^0)\cosh(H\lVert\overrightarrow{X}\lVert)\right),$$
 where ${\lVert\overrightarrow{X}\lVert}=(X_1^2+X_2^2+X_3^2)^{\frac{1}{2}}.$ 
By using this coordinate system, it is easy to show that the dS flat waves of massive fields $\left(x\cdot\xi\right)^{-2\pm i\nu}$, $\left(x\cdot\xi\right)^{-1\pm i\nu}$ or $\left(x\cdot\xi\right)^{\frac{3}{2}\pm i\nu}$, in null curvature limit reduce to Minkowski flat wave $e^{\pm ik\cdot X}$ \cite{ta97}. Also for massless vector fields, the dS flat wave $\left(x\cdot\xi\right)^{-2}$ or $\left(x\cdot\xi\right)^{-1}$ can be mapped to intrinsic coordinate and then it can be shown that the null curvature limit of them match with their Minkowskian counterparts \cite{gagarota,ta97}.
Also, one can show that \cite{ta97}:
 $$
 \lim_{H\rightarrow 0}H\cancel{x}=\lim_{H\rightarrow 0}H\eta_{\alpha\beta}\gamma^\alpha x^\beta=-\gamma^4\,.
 $$
 For dS massless vector field one can show that the polarization states are \cite{gagarota,ta97}:
 $$\lim_{H\rightarrow 0}{\cal E}_{2\alpha}(x,\xi,n)=  \lim_{H\rightarrow 0}{\cal E}_{1\alpha}(x,\xi,n)=\epsilon_\mu^{(n)}\, ,$$
 where
 $\mu=0,1,2,3$ and $\epsilon_\mu^{(n)}$ is the polarization of massless vector field in Minkowski space-time.
 By attention to Dirac equation for spinor \eqref{ds sp fi eq} and adjoint spinor \eqref{ad ds sp fi eq} one can see that the null curvature limit of 
$\Psi$
and
$\bar{\Psi}$ are:
\b\label{psi flat limit}
\lim_{H\rightarrow 0}\Psi(x)=\psi(X)
\;,\;\;\;
\lim_{H\rightarrow 0}\bar{\Psi}(x)\gamma^4=\bar{\psi}(X).
\e
Thus by attention to \eqref{psi flat limit} and \eqref{ds spinor two-point function} one can obtain:

\b\label{sp-fi 2point function flat limit}
\lim_{H \rightarrow 0}{\bf S}(x_1,x_2)=
-i\lim_{H \rightarrow 0}\le\Omega\lvert \Psi(x_1)\;\bar{\Psi}(x_2)\lvert\Omega\re=
i\le 0\lvert \;\psi(X_1)\; \bar{\psi}(X_2)\gamma^4\;\lvert 0\re
=-S(X_1,X_2)\gamma^4.
\e
Therefore one can obtain the null curvature limit of $\chi^{dS}$ as:
$$
\begin{array}{cl}
&\displaystyle\lim_{H \rightarrow 0}{\chi}^{dS}  
=-\dfrac{iq^3}{\hslash^3}
{\cal N}_k{\cal N}_{k'}{\cal N}_{k''}
\displaystyle\int\int\int \ud^4X_1 \ud^4X_2 \ud^4X_3
e^{-ik\cdot X_1} e^{-ik'\cdot X_2} e^{-ik''\cdot X_3}
\\
\\
&
\times\, Tr
\left\{
\gamma^4\left(-\gamma^4\right)\epsilon_\mu\gamma^\mu
\left(S(X_1,X_2)\gamma^4\right)
\gamma^4\left(-\gamma^4\right)\epsilon_\nu\gamma_\nu
\left(S(X_2,X_3)\gamma^4\right)
\gamma^4\left(-\gamma^4\right)\epsilon_\lambda\gamma^\lambda
\left(S(X_3,X_1)\gamma^4\right)
\right\},
\end{array}
$$
By using \eqref{minkofski gamma} and Fourier transformation of spinor two-point function $S(X,X')$, and also by attention to that $\gamma^4\gamma^4=-I$, one can obtain:
$$
\begin{array}{clcr}
\displaystyle\lim_{H\rightarrow 0}\chi^{dS}&=\displaystyle\dfrac{1}{(2\pi)^{12}}\dfrac{iq^3}{\hslash^3}
{\cal N}_k{\cal N}_{k'}{\cal N}_{k''}
\int\int\int \ud^4p\;\ud^4p' \ud^4p''
\dfrac{
	Tr\left[
	\cancel{\epsilon}\left(\barra{p}+m\right)\cancel{\epsilon}^{\;'}\left(\barra{p}'+m\right)\cancel{\epsilon}^{\;''}\left(\barra{p}''+m\right)
	\right]
}
{
	\left(p^2-m^2+i\tau\right)\left(p'^2-m^2+i\tau\right)\left(p''^2-m^2+i\tau\right)
}\;
\\
\\
&\displaystyle\times\int\int\int \ud^4X_1 \ud^4X_2 \ud^4X_3
\left\{
e^{-ik\cdot X_1}
e^{ik'\cdot X_2}
e^{ik''\cdot X_3}
e^{-ip\cdot (X_1-X_2)}
e^{-ip'\cdot (X_2-X_3)}
e^{-ip''\cdot (X_3-X_1)}
\right\},
\end{array}
$$
where, $\barra\epsilon=\epsilon_\mu\gamma^{'\mu}\;.$ After some mathematical calculations the $\displaystyle\lim_{H\rightarrow 0}\chi^{dS}$ is  obtained as:
$$
\begin{array}{clcr}
\displaystyle\lim_{H\rightarrow 0}\chi^{dS}
=\displaystyle\dfrac{iq^3}{\hslash^3}
{\cal N}_k{\cal N}_{k'}{\cal N}_{k''} \delta^4(k-k'-k'')
\\
\\
\displaystyle
\times\int\ud^4p''\;\dfrac{
	Tr\left[
	\left(\barra{p}+m\right)
	\cancel{\epsilon}
	\left(\barra{p}-\barra{k}+m\right)
	\cancel{\epsilon}^{\;'}
	\left(\barra{p}-\barra{k}''+m\right)
	\cancel{\epsilon}^{\;''}
	\right]
}
{
	\left(p^2-m^2+i\tau\right)
	\left((p-k)^2-m^2+i\tau\right)
	\left((p-k'')^2-m^2+i\tau\tau\right)
}\;.
\end{array}
$$
As it seen, the null curvature limit of ${\cal S}^{(3)}_{\gamma\rightarrow\gamma'+\gamma''}$ is matched on Minkowski result \eqref{xi} exactly.
\\
\setcounter{equation}{0}
\section{Conclusion}\label{conclusion}

A photon decaying to two photons in dS ambient space formalism is investigated and its ${\cal S}$ matrix elememts are obtained in the first approximation. Indirect gravitational effect appears in this ${\cal S}$ matrix elememts from the two-point functions and the polarization vectors. This gravitational effect can be calculated explicitly by numerical methods which maybe done in the next work. It was shown that the ${\cal S}$ matrix elememts in the null curvature limit exactly reduce to the Minkowskian counterparts.

The Hamiltonian interaction of three photons was obtained from quantum field theory approach. 
The interaction constant $\chi$ is approximately obtained theoretically. This constant is depend on the energy-momentum and the polarization states of photons. These photons interact with electrons of the spinor loop or virtual electrons which appear in the quantum vacuum fluctuations. It should be noted that these electrons are not electrons of the matter or crystal, these are electrons that create and annihilate in quantum vacuum fluctuations that interact with the real photons. These results are obtained for one incoming photon and we ignore the bulk effects. The bulk effects could be investigated in the next works.

\vspace{0.5cm} 
\noindent{\bf{Acknowledgements}}:  The authors wishes to express their particular thanks to A. Rabeei, M. Amiri, M. Rastiveis and R. Razani for useful discussions.


\end{document}